\documentclass[submission, PhysProc]{SciPost}

\hypersetup{
    colorlinks,
    linkcolor={red!50!black},
    citecolor={blue!50!black},
    urlcolor={blue!80!black}
}

\usepackage[bitstream-charter]{mathdesign}
\DeclareSymbolFont{usualmathcal}{OMS}{cmsy}{m}{n}
\DeclareSymbolFontAlphabet{\mathcal}{usualmathcal}

\begin{document}

\begin{center}{\Large \textbf{How to determine the shape of nuclear molecules with polarized gamma-rays
\\
}}\end{center}

\begin{center}
Lorenzo Fortunato\textsuperscript{1,2,*}
\end{center}

\begin{center}
{\bf 1} Dip. Fisica e Astronomia "G. Galilei", Univ. Padova, Padova, Italy
\\
{\bf 2} I.N.F.N. Sez. di Padova, via Marzolo, 8, I-35131 Padova, Italy
\\
${}^\star$ {\small \sf fortunat@pd.infn.it}
\end{center}

\begin{center}
\today
\end{center}


\definecolor{palegray}{gray}{0.95}
\begin{center}
\colorbox{palegray}{
  \begin{tabular}{rr}
  \begin{minipage}{0.05\textwidth}
    \includegraphics[width=14mm]{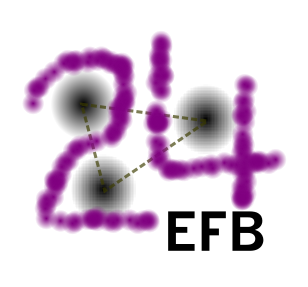}
  \end{minipage}
  &
  \begin{minipage}{0.82\textwidth}
    \begin{center}
    {\it Proceedings for the 24th edition of European Few Body Conference,}\\
    {\it Surrey, UK, 2-4 September 2019} \\
    \end{center}
  \end{minipage}
\end{tabular}
}
\end{center}

\section*{Abstract}
{\bf
A method has been recently proposed to establish the geometry of
the alpha-cluster arrangement in $^{12}$C making use of polarized gamma-rays. The ratio of intensities of
scattered radiation at 90 degrees along and perpendicular to the initial direction of the electric field
vector, called depolarization ratio, is a key quantity that allows to underpin the nature of totally
symmetric modes of vibrations. This allows to connect with the underlying point-group structure
and therefore to the geometric shape of the nuclear molecule.
This method is reviewed for $^{12}$C and extended to other configurations, such as three unequal clusters and four identical clusters (e.g. $^{16}$O).
}

\vspace{10pt}
\noindent\rule{\textwidth}{1pt}
\tableofcontents\thispagestyle{fancy}
\noindent\rule{\textwidth}{1pt}
\vspace{10pt}

\section{Introduction}
\label{sec:intro}
Chemical molecules are rather rigid bound quantum systems of atoms that perform vibrations and rotations around fixed equilibrium positions and the Bohr-Oppenheimer approximation (i.e. decoupling of electronic and nuclear motion) is valid in this context. Their geometry has been investigated with Raman scattering, by shining electromagnetic radiation (in various frequency domains: microwaves, infrared, visible, etc.) and observing the scattered radiation at different angles. The response of the molecule to the electromagnetic field contains information on the geometry of the molecule itself that are encoded in the measured spectra through the polarizability tensor, $\alpha$. This tensor is related to the response of a system of charges to an external electric field and the polarizability vector is related to it as 
\begin{equation}
\vec P = \alpha \vec E
\end{equation}
If the incoming photon beam is polarized (for the rest of the paper by polarized we mean linearly polarized, i.e. the polarization plane is fixed in space and the direction of the electric field vector is constant), the outcoming radiation intensity can be passed through a polarimeter that acts as an analyzer (See Fig. \ref{ram}). Usually the scattered intensity is measured along the direction perpendicular to the incoming beam (although it is not mandatory), therefore for the rest of the paper we will refer to that case.
The peaks in the Raman spectra can be measured with the polarimeter aligned along the initial direction of polarization and along a perpendicular direction and a ratio of these two quantities can be derived, that is usually called depolarization ratio, defined as:
\begin{equation}
\rho =\frac{I_\perp}{I_\parallel} \;.
\end{equation}
This quantity is connected to the asymmetry, that is more commonly used in nuclear physics\cite{Asymm}, via the formulas: 
\begin{equation}
A=\frac{I_\parallel-I_\perp}{I_\parallel + I_\perp}= \frac{1-\rho}{1+\rho} \;.
\end{equation}
Several other definitions are in use that depend on the angle of scattering, but in general they are all related.
In quantum mechanics, precise relations for the intensities of light scattered from a random sample can be found that are due to Placzek \cite{Placzek,Long} through the so-called theory of classical radiation scattering that makes use of the Placzek invariants. Intensities and ratios thereof can be calculated analytically and lead to simple expressions. In particular, the value of $\rho$ becomes:
\begin{equation}
\rho =\frac{3\gamma^2}{45a^2+4\gamma^2} \;,
\end{equation}
where $a$, the mean polarizability, and $\gamma$, the anisotropy, are related to Placzek invariants as follows:
\begin{equation}
    3a^2= {\cal G}^{(0)} \qquad 2\gamma^2 =3{\cal G}^{(2)}
\end{equation}
and these are connected to components of the polarizability tensor, $\alpha_{ij}$ with $i,j=\{x,y,z\}$ as follows:
\begin{align}
    {\cal G}^{(0)} &= \frac{1}{3}|\alpha_{xx}+\alpha_{yy}+\alpha_{zz}|^2 \\
     {\cal G}^{(2)} &= \frac{1}{2}\Bigl(|\alpha_{xy}+\alpha_{yx}|^2+|\alpha_{xz}+\alpha_{zx}|^2+|\alpha_{yz}+\alpha_{zy}|^2\Bigr)+\nonumber\\ &+\frac{1}{3}\Bigl(|\alpha_{xx}-\alpha_{yy}|^2+|\alpha_{xx}-\alpha_{zz}|^2+|\alpha_{yy}-\alpha_{zz}|^2\Bigr)
\end{align}
Many other definitions are used in different books of quantum chemistry and optics, therefore extra care should be taken when considering these quantities. 
Analogous expressions can be found for general scattering angles, that are not relevant for the scopes of the present paper (Ref. \cite{Long} reports the most common in Appendix).
Under certain circumstances, the ratio above goes exactly to $3/4$ and this forms a sort of a benchmark. For all the Raman peaks that belong to the totally symmetric representation of a given group, the following is true:
\begin{equation}
0\le \rho \le 3/4
\end{equation}
while for all the others, i.e. those belonging to non-symmetry representations, the following theorem holds:
\begin{equation}
\rho = 3/4
\end{equation}
This is duly observed in chemistry, where it has been used to probe the geometric structure of several molecules (Example: CCl$_4$ as discussed during the presentation).

\section{Clustering in nuclei and nuclear molecules}
It is well known that clusters, and especially alpha clusters, play a fundamental role in the nuclear structure of light nuclei (for a review see \cite{Beck}). For example, $^8$Be only exists as a resonant $L=0$ state of two alpha particles that can even rotate up to $L=4$ around a common center of mass. Nuclei and the nucleons that move inside them are most certainly not rigid systems as the chemical molecules, nonetheless it is very profitable to apply the language and the methods of quantum chemistry to nuclear systems, because some light nuclei, or maybe only a part of their spectrum, can be described as a sort of soft nuclear molecule, in which the clusters (that take the place of the atoms) perform large amplitude fluctuations around the equilibrium points, the mean value of kinetic energy of these subsystem being very large and comparable with typical nuclear excitation energies.

Without entering into the details, one should mention that algebraic models have been proposed for $\alpha-$clustered nuclei. Notably, Iachello and Bijker have introduced the algebraic cluster model \cite{Bij,Bij2} that has been very successful in explaining the low-energy spectrum of $^{12}$C and $^{16}$O in terms of vibrations and rotations of equilateral and tetrahedral cluster structures respectively \cite{Bij,Bij4}. These studies have generated a large experimental interest \cite{Fre, Gai}, of which is impossible to give an account in these pages (see also \cite{Beck}).

Other interesting algebraic schemes have also been proposed, such as the Semimicroscopic Algebraic Cluster Model \cite{Cseh, Hess}, that incorporates the effects of the Pauli exclusion principle.

I will examine, in the following, the case of three alpha particles, by listing all of the possible molecular shapes that this partition of the mass of $^{12}$C can take. I will determine the point-group symmetry and the character of vibrational modes of motion and I will relate this information to the expected outcome of an experiment of Raman spectroscopy. 

\begin{figure}[!t]\begin{center}
	\includegraphics[width=.8\columnwidth, clip=]{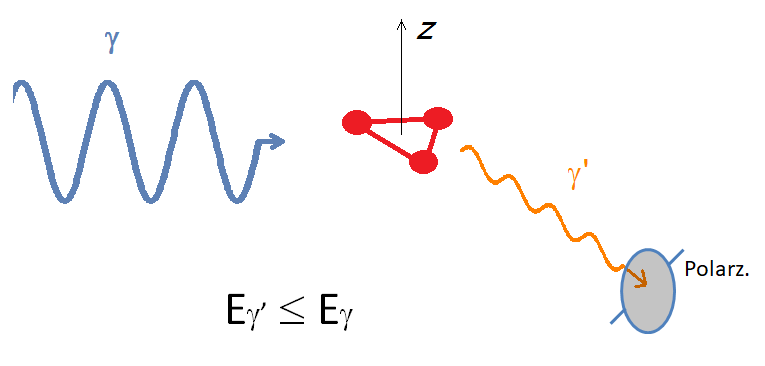}
	\end{center}
	\caption{Schematic representation of the scattering of polarized gamma rays at 90$^o$ from a sample. The outgoing radiation is analyzed with a polarimeter.}
	\label{ram}
\end{figure}

Ideally, one should realize a Raman experiment like the one schematically depicted in Fig. \ref{ram}, in which linearly polarized gamma radiation is sent to a sample and the light emitted at 90$^o$ is measured through a polarizer. The resulting ratio of perpendicular and parallel intensities can be correlated with the appearance of totally symmetric representations of a given group.
In fact, due to the theory cited above, a depolarization ratio of three quarters will signal non-totally symmetric modes (of any kind), while a ratio significantly different from 3/4 will instead signal a totally symmetric mode.

\section{Three clusters}
\label{sec:three_clusters}
\begin{figure}[!t]
	\includegraphics[width=1.\columnwidth, clip=]{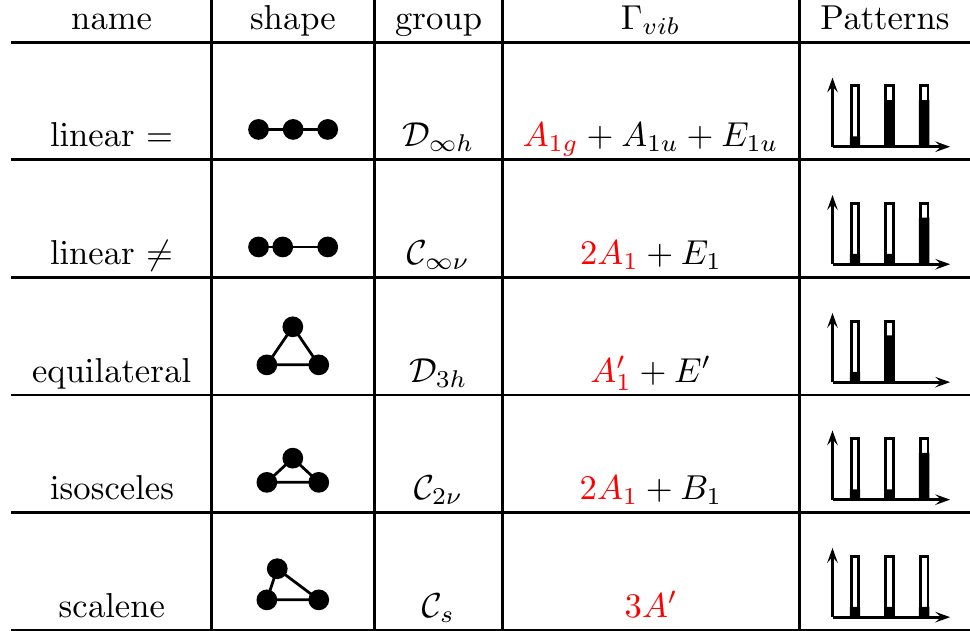}
	\caption{This table enumerates the possible shapes of a nuclear molecule made out of three identical clusters, their underlying point-group symmetry, and the characters of the normal modes of vibration (with totally symmetric modes highlighted). The last column shows the schematic predicted patterns for depolarization ratios (Intensity vs. Energy, not to scale). The white/black histograms represent parallel/perpendicular intensity.}
	\label{Tab1}
\end{figure}

Three equal clusters (of bosonic type for the present paper) can take only 5 different spatial configurations. They can be on a line at equal distances from each other or at different distances or they can be out of a straight line (thus lying on a plane) forming a triangle that can be equilateral, isosceles or scalene. There are no other possibilities. The so-called bent configuration is nothing but an isosceles triangle if it has equal arms and a scalene triangle if it has unequal arms.
Each of these shapes corresponds to a precise discrete point-group symmetry that one can find in Table \ref{Tab1}. This is easily determined through standard group theoretical methods (cfr. Ref. \cite{Carter}). These groups are the linear centrosymmetric group, ${\cal D}_{\infty h}$, the linear non-centrosymmetric group, ${\cal C}_{\infty \nu}$, the full dihedral group of order 3, ${\cal D}_{3h}$, the cyclic pyramidal symmetry group of order 2, ${\cal C}_{2\nu}$ and the reflection symmetry group, ${\cal C}_{s}$.

The system with $N=3$ particles has a total of $3N=9$ degrees of freedom and upon subtraction of 3 d.o.f for translations and 2 or 3 (depending if the molecule is linear or not) for rotations, one ends up with $3N-5=4$ or $3N-6=3$ vibrational degrees of freedom for linear and planar configurations respectively. These might be singly-degenerate representations of a group (indicated with $A$ or $B$)  of doubly-degenerate representations (indicated with $E$)\footnote{Or even triply-degenerate (indicated with $T$ or $F$), but not in the present case of three identical clusters}.
Now, the fourth column of Tab. 1 shows the number and type of vibrational characters of each configuration, $\Gamma_{vib}=\Gamma_{3N}-\Gamma_{tras}-\Gamma_{rot}$, and the totally symmetric are highlighted in red. Notice that doubly-degenerate characters have been comprised in a single peak in the last column. These schematic pictures should be read as follows: all the vibrational states plus all the rotational states built on top of them (i.e. the whole rotational band, indicated with the histograms) with totally symmetric characters will show a perpendicular intensity (black color) that sits in between 0 and 3/4 of the parallel intensity (white color), while all the other will show exactly 3/4. This is a very robust signature.
Notice that, with the exception of the second and fourth predictions, all the others are mutually exclusive. If a measurement could be done, while scanning in energy with the gamma rays, a strong indication of the characters of each state could be reached. Ideally, by this method, it should be possible to attach a label "totally symmetric" or "non-totally symmetric" to each feature of the spectrum, and this information, combined with decays and theoretical interpretation should give an answer to the question of the geometrical arrangements.

There are several factors affecting the successful outcome of this proposed experiment. The most important is the fact that, in nuclei, many of the states of interest lie above a separation threshold. In $^{12}$C, only the g.s, and the first excited $2^+$ state have an energy lower than the separation into $3\alpha$ particles, therefore many higher lying states would prefer particle decay to gamma decay. For example, the Hoyle state decays with branching ratio > 99.9$\%$ into the $^8$Be+$^4$He channel, while the direct $3\alpha$ decays amount to about <0.04$\%$, leaving little space to the $2\gamma$ decay (about 0.04$\%$), according to recent measurements \cite{meas}.

\begin{figure}[!t]\begin{center}
	\includegraphics[width=0.4\columnwidth, clip=]{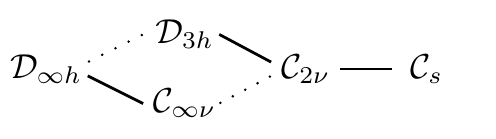}
	\caption{Chains of group-subgroup relations for the present case.}
	\label{cha}\end{center}
\end{figure}

\begin{figure}[!t]
	\includegraphics[width=1.\columnwidth, clip=]{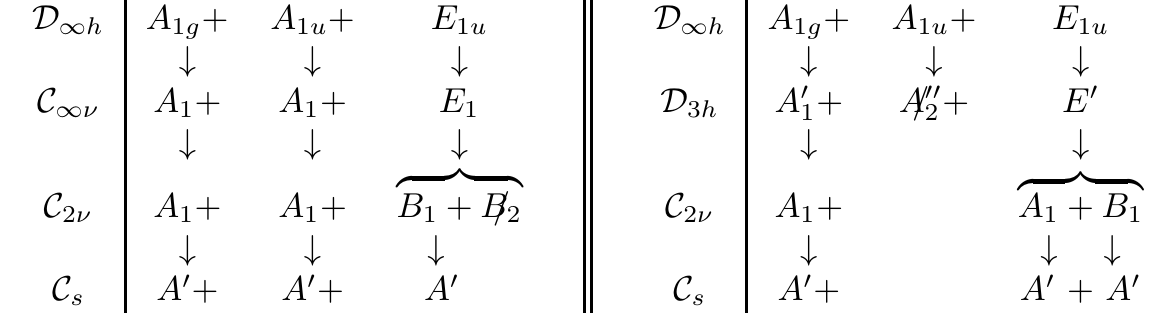}
	\caption{Descent in symmetry: connections between the representations of the group and those of its subgroups.}
	\label{des}
\end{figure}

One can wonder also what happens if one state (for instance the g.s.) has one symmetry and another has a different symmetry. In that case, the lower symmetry should be used as a guidance and there are ways (i.e. the procedure of descent in symmetry) to relate the character of a certain representation of the higher group with the character of representations of the lower group, see Fig. \ref{cha} and \ref{des}. Very often the totally symmetric modes are linked and therefore are easy to spot.

\section{Some extensions}
\begin{figure}[!t]\begin{center}
	\includegraphics[width=.8\columnwidth, clip=]{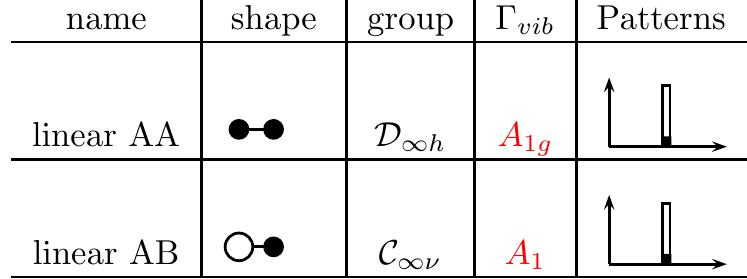}
	\caption{Nuclear molecules made out of two identical or different clusters, their underlying point-group symmetry, and the characters of the only normal mode of vibration (that is always totally symmetric).}
	\label{Tab2}\end{center}
\end{figure}

This concept can be easily adapted to other cases. The case of $N=2$ clusters, either equal to each other or different, is not very interesting as the only possible vibration is always of totally symmetric character, as in Fig. \ref{Tab2}. In this case, the Raman fluorescence experiment cannot be used to discriminate among various arrangements, but anyway we might use it as a test for clusterization, for instance I would expect the depolarization ratio to be somewhere between 0 and 3/4 for the $A=7$ isobars, like $^7$Li and $^7$Be, that are known to possess a pronounced $\alpha+t$ or $\alpha+h$ cluster structure.

\begin{figure}[!t]
	\includegraphics[width=1.\columnwidth, clip=]{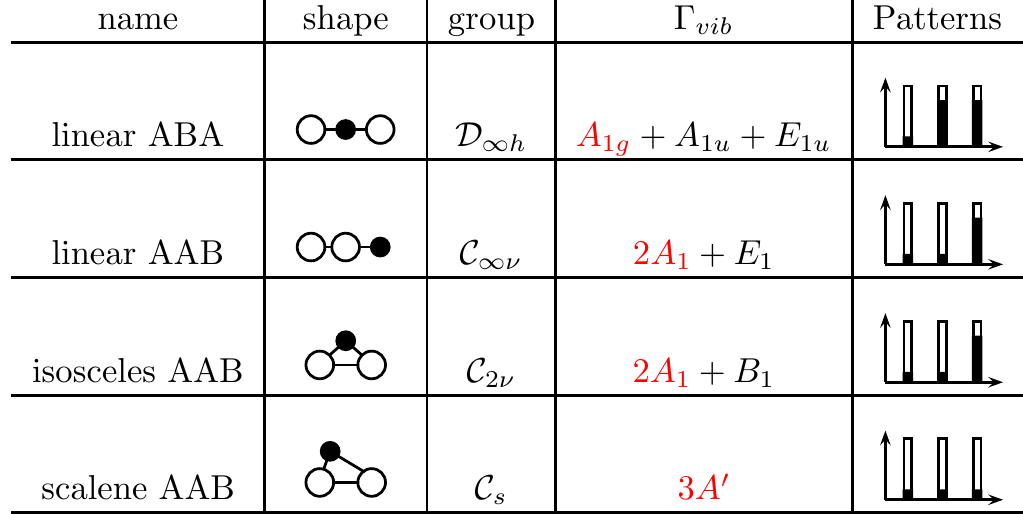}
	\caption{Configurations of three clusters, one of which is different from the other two, their underlying point-group symmetry, and the characters of the only normal mode of vibration.}
	\label{Tab3}
\end{figure}

Similarly one might wonder on how to extend the present model to three clusters, one of which is different from the other two. In Fig. \ref{Tab3} we show all the possible outcomes.
Notice that one might naively expect other configurations, but they are equivalent to those listed in the table. For example, there could be a linear configuration of type ABA, but with unequal arms' lengths. This is equivalent to AAB, i.e. it is non-centrosymmetric.  Another case is the equilateral triangle, here is absent because its symmetry is of the same type of the isosceles triangle. For the sake of illustrating this point further: if one rotates an equilateral triangle with one vertex different from the others of 120$^o$ with respect to the axis perpendicular to the plane of the molecule, the resulting figure differs from the initial one, therefore it is not invariant with respect to all the operations of the ${\cal D}_{3h}$ group, but only with respect to those of the lower group ${\cal C}_{2\nu}$.
One might devise experiments to measure the vibrational states and the rotational bands built on top of them also in systems like $^{11}$C or $^{11}$B and try to see if they fit the predictions of any of the arrangements listed in the table. Notice that the second and third case give the same patterns. 

The same scheme can be devised in the case of $^{16}$O and I have already undertaken several calculations in this respect, but the number of possible linear, planar or spatial configurations of four alpha particles with some special geometric symmetry is quite big (for instance square, kite, parallelogram, tetrahedron, elongated or compressed tetrahedron, and so on...) and several of them show the same or very-similar patterns, therefore this method loses some of its appeal for $N\ge4$. 

\section{Other developments: modeling reactions in $^{12}$C}
\begin{figure}[!t]\begin{center}
	\includegraphics[width=0.4\columnwidth, clip=]{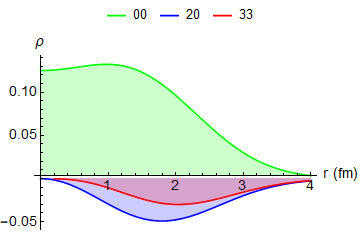}
		\includegraphics[width=0.4\columnwidth, clip=]{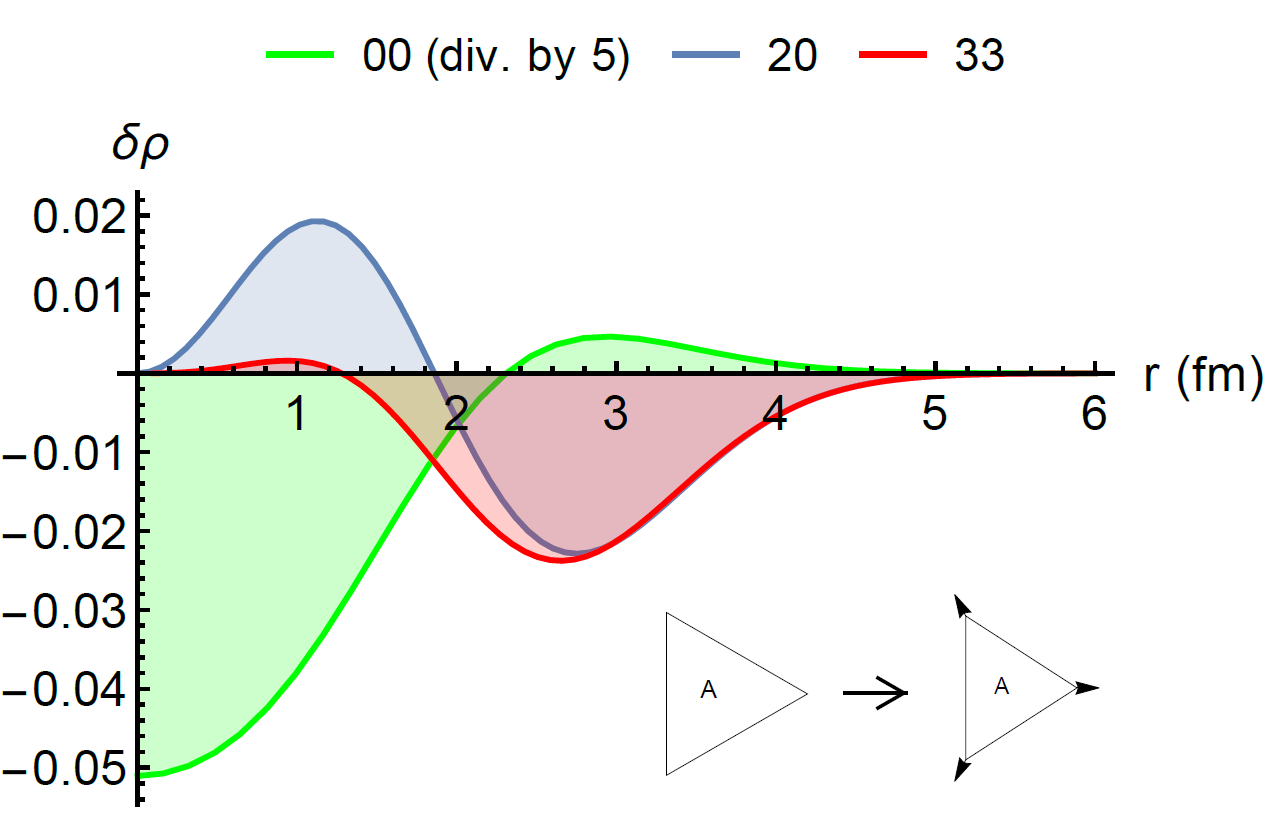}\end{center}
	\caption{Left: Radial components of the expansion in spherical harmonics of the g.s. density distribution for $\lambda\mu=00,20$ and $33$. Right: transition densities from the ground band to the Hoyle band, that is a breathing vibrations with character A.}
	\label{dens}
\end{figure}

Very recently, we have started looking at the densities and transition densities of $^{12}$C in the cluster model using gaussian density distributions for the alpha particles in a way similar to that used in Ref. \cite{Del17b}. The total density is made up from three alpha particles at the vertexes of a triangle. From densities, we calculate electromagnetic transition probabilities, nuclear folding potentials and form factors that can be used to model elastic and inelastic scattering processes such as $^{12}$C + $\alpha$.
For example, in Fig. \ref{dens} we show in the left part the radial components of the expansion in spherical harmonics of the g.s. density distribution for $\lambda\mu=00,20$ and $33$, i.e. the first terms of 
\begin{equation}
\rho_{gs}(\vec r)= \sum_{\lambda\mu} \rho_{gs}^{\lambda,\mu} (r) Y_{\lambda,\mu} (\theta,\varphi)
\end{equation}
and, in the right part of the picture, we show the transition densities from the ground state band to the Hoyle band, 
\begin{equation}
\delta \rho_{gs \rightarrow A} (\vec r) =\sum_{\lambda\mu}\delta \rho_{gs \rightarrow A}^{\lambda\mu}(r)Y_{\lambda \mu} (\theta,\varphi)
\label{expdrho}
\end{equation}
where 
\begin{equation}
\delta \rho_{gs \rightarrow A} (\vec r) =\chi_1\frac{d}{d\beta}\rho_{gs}(\vec r, \beta)~.
\label{chi1}
\end{equation}
and where $\chi_1$ is the intrinsic transition matrix element. 
One can notice that the $00$ term is the most relevant and this will give a significant E0 moment. A complete study on all the densities and transition densities within the ground state band, the excited A-type vibrational band (Hoyle band) and the excited E-type vibrational band, and the intra-band transitions will form the subject of a future paper to which we are currently working with the aim of demonstrating that the simple triangular model is able to satisfactorily describe the features of the inelastic scattering of alpha particles on $^{12}$C.

\section{Conclusions} \label{sec:conclusions}
The method discussed in the present contribution can give crucial information for the determination of a nuclear cluster configuration. An experiment of Raman spectroscopy with polarized photons in the range of typical nuclear energies (0-20 MeV) would lead to a scheme of depolarization ratios that can be tested against the various possibilities enumerated in the tables, thus allowing to resolve the riddle of the cluster configurations of $^{12}$C and other nuclei, or at least to restrict the list of possible spatial arrangements.




\begin{thebibliography}{99}
\bibitem{2019FO} L. Fortunato, {\it Establishing the geometry of $\alpha$ clusters in $^{12}$C through patterns of polarized $\gamma$ rays}, Phys. Rev. C {\bf 99}, 031302(R) (2019), \doi{10.1103/PhysRevC.99.031302}.
\bibitem{Asymm} L.W. Fagg and S.S. Hanna,  Rev.Mod.Phys. {\bf 31}, n.3 (1959)
\bibitem{Long} D.A. Long, {\it  The Raman Effect: A Unified Treatment of the Theory of Raman Scattering by Molecules.} (2002) John Wiley \& Sons Ltd, U.K.
\bibitem{Placzek} Placzek, G. (1934), {\it Rayleigh-Streuung und Raman-Effekt}, in {\it Handbuch der Radiologie, (ed. E. Marx)}, 6, 205–374. Academische Verlag: Leipzig.
\bibitem{Beck} C. Beck, {Clusters in Nuclei, Vol. I, II and III}, Lecture Notes in Physics {\bf 818} (2010), ibidem {\bf 848} (2012),  ibidem {\bf 875} (2014)
\bibitem{Bij} R. Bijker and F. Iachello, {\it The algebraic cluster model. Three-body clusters}, Annals Phys. {\bf 298} (2002) 334-360
\bibitem{Bij2} R. Bijker,{\it Algebraic treatment of alpha-cluster nuclei}, J. Phys.: Conf. Ser.{\bf 492}, 012009 (2014)
\bibitem{Bij4} R. Bijker and F. Iachello, {\it The Algebraic Cluster Model: Structure of 16O} Nucl. Phys. A {\bf 957}, 154-176 (2017) 
\bibitem{Fre} D.J, Marin-Lambarri et al., {\it Evidence for Triangular D3h Symmetry in 12C}, Phys.Rev.Lett. {\bf 113} (2014) no.1, 012502 
\bibitem{Gai} M. Gai, {\it The ${\cal D}_{3h}$ symmetry of 12C}, AIP Conference Proceedings 2150, 030007 (2019)
\bibitem{Cseh}J. Cseh, G. Levai, A. Algora, P. O. Hess, K. Kato, {\it The Semimicroscopic Algebraic Cluster Model: I. — Basic concepts and relations to other models}, Il Nuovo Cimento A (1971-1996) (1997) 110: 921
\bibitem{Hess} P.O. Hess, J.R.M. Berriel-Aguayo,L.J. Ch\'avez-Nu{\~n}ez, {\it 16O within the Semimicroscopic Algebraic Cluster Model and the importance of the Pauli Exclusion Principle}, Eur. Phys. J. A (2019) 55: 71
\bibitem{Carter} R.L. Carter, {\it Molecular symmetry and group theory},  Wiley-India (2012)
\bibitem{meas}  B. Alshahrani, T. Kibédi, A.E. Stuchbery, E. Williams and S. Fares,  EPJ Web of Conferences {\bf 63}  01022 (2013), Heavy Ion Accelerator Symposium 2013.
\bibitem{AIP} A. Vitturi, J. Casal, L. Fortunato, and E.G. Lanza, AIP Conference Proceedings 2150, 040006 (2019), \doi{10.1063/1.5124607}.
\bibitem{Del17b} V. Della Rocca, R. Bijker, F. Iachello, {\it Single-particle levels in cluster potentials}, Nucl. Phys. A{\bf 966} (2017) 158-184, \doi{10.1016/j.nuclphysa.2017.06.032}

\end{thebibliography}

\nolinenumbers

\end{document}